\documentclass{aa}     
\usepackage{txfonts} 
\usepackage{graphicx}    
      
\begin{document}     
     
\title{Cosmic Microwave Background Polarization: constraining
models with a double reionization}     

\author{L.P.L.~Colombo\inst{1,2} \and G.~Bernardi\inst{3} \and 
L.~Casarini\inst{3} \and R.~Mainini\inst{1,2} \and S.A.~Bonometto\inst{1,2}
\and E.~Carretti\inst{3} \and R.~Fabbri\inst{4}}
     
\institute{Physics department `G.~Occhialini', University of
Milano--Bicocca, Piazza della Scienza 3, I20126 Milano, Italy \and
INFN, Sezione di Milano, Via Celoria 16, I20133 Milano, Italy \and
I.A.S.F./C.N.R. Bologna, Via Gobetti 101, I-40129 Bologna, Italy \and
Physics department, University of Firenze, Via Sansone 1,
I-50019 Sesto Fiorentino, Italy}     

\offprints{L.P.L. Colombo, \email{loris.colombo@mib.infn.it}}

\titlerunning{CMBP and reionization: constraining
models with a double reionization}

\authorrunning{L.P.L. Colombo et al.}

\date{Received  /Accepted }     
     
\abstract {Neutral hydrogen around high--$z$ QSO and an optical depth
$\tau \sim 0.17$ can be reconciled if reionization is more complex
than a single transition at $z\simeq 6-8$. Tracing its details could
shed a new light on the first sources of radiation. Here we discuss
how far such details can be inspected through planned experiments on
CMB large-scale anisotropy and polarization, by simulating an actual
data analysis.  By considering a set of double reionization histories
of Cen (\cite{cen}) type, a relevant class of models not yet
considered by previous works, we confirm that large angle experiments
rival high resolution ones in reconstructing the reionization history.
We also confirm that reionization histories, studied with the prior of
a single and sharp reionization, yield a biased $\tau $, showing that
this bias is generic. We further find a monotonic trend in the bias
for the models that we consider, and propose an explanation of the
trend, as well as the overall bias. We also show that in long-lived
experiments such a trend can be used to discriminate between single
and double reionization patterns.

}
     
\maketitle     

\section{Introduction}

\label{sec:intro}

The first--year WMAP data release\footnote{
http://lambda.gsfc.nasa.gov/product/map/} detected a strong
anisotropy--polarization cross--correlation at low $l$ (Ko\-g\-ut et
al.~ \cite{wmap:kogut}). Previous analyses suggested that the
Intergalactic Medium (IGM) had reionized at a redshift $z\simeq 6$--8,
and led one to expect an optical depth for Thomson-scattering $\tau
\la 0.05$ (Miralda--Escud\`e~ \cite{mira}). Observations of the
Gunn-Peterson effect in high--$z$ QSO, requiring a fraction of neutral
hydrogen at $z\sim 6$--7, agreed with this scheme (Djorgovski et
al.~\cite{djor}, Becker et al.~\cite{becker}; see however Malhotra \&
Rhoads~\cite{malho}). On the contrary, the level of
anisotropy--polarization correlation, in WMAP data, indicates $\tau
\approx 0.17$ (Kogut et al.~\cite{wmap:kogut}) and reionization at
$z\ga 16$, assuming a single-step reionization model.  If so, the
reionization history could have been quite complex. In turn, data on
CMB anisotropy and polarization could shed new light on the birth and
evolution of primeval objects

Different options on the nature of primeval objects have been
considered. Ciardi et al. (\cite{ciardi}) showed that metal--free
stars in early galaxies may account for a depth up to $\tau \approx
0.15$, although this seems an upper limit to such a picture (Ricotti
\& Ostriker \cite{ro04a}). Various authors also considered an early
pre-ionization due to black holes in small galaxies (Ricotti \&
Ostriker \cite{ro04b}) or miniquasars (Madau et al. \cite{madau}), or
even the effects of sterile neutrino decay (Hansen \& Haiman \cite{haha}).

Independently of the ionizing mechanism, however, comparing WMAP and
QSO data suggests that reionization is not achieved in a single, rapid
step, but involves at least two different stages. Double reionization
models were suggested, even before WMAP, by Cen (\cite{cen}), Wyithe
et al. (\cite{wsl} ), Sokasian et al. (\cite{soka}), Ricotti \&
Ostriker (\cite{ro04a}) and others. In these models the ionization
fraction attains a value $x_e \approx 1$, at some high $z$, then
partial recombination occurs, while a second reionization takes place
at $z=6$--8 and then $x_e=1$ (for discussion purposes, we neglect
Helium ionization).

Haiman \& Holder (\cite{haho}) also considered another option, that
the Universe partially reionized (up to $x_e \sim 0.5$--0.8) at a high
$z_r$, to achieve a complete reionization at $z \simeq 6$--8. A
similar \textit{ 2--step reionization} was also schematically
considered by Kaplinghat et al.  \cite{kap}, for its effects on
CMB. Still different reionization patterns were treated by Bruscoli et
al. (\cite{bruscoli}), Hu \& Holder (\cite{huho} ), Naselsky \& Chiang
(\cite{nachiang}) and Colombo (\cite{colombo}), confirming the
interest in the relation between CMB data and early reionization,
although often specifying no link with early object formation models.
The present work is based on the dependence of the low--$l$ behavior
of the $C_l^E$ spectra on the whole reionization history, and aims to
predict the actual detectability of $\tau $, as well as of the
reionization redshift(s), the ionization rate(s), etc., with data on
large angular scales.  When dealing with full--sky small--angle
experiments, the high number of pixels forces one to work in harmonic
space which, under the assumption of a Gaussian signal, allows a fair
information compression. For large--angle data, working directly in
pixel space, as opposed to harmonic space, is numerically feasible and
allows one to take into account a number of features that experiments
can hardly avoid.  

In particular, the actual CMB signal can be
recovered only on a portion of the celestial sphere. An example of
such a limitation is the effect of Galactic contamination, making
anisotropy data unreliable within $\sim 20 \degr$ of the galactic
plane.  On the other hand, first year WMAP data indicate that the main
polarized foreground at frequencies up to 70~GHz is Galactic
synchrotron (Bennett et al. \cite{wmap:bennett}), whose polarization
spectrum has a steep dependency on frequency.  According to the
synchrotron template by Bernardi et al. (\cite{bernardi}), a lesser
contamination is expected in polarization and the analysis can avoid
Galactic cuts for such a signal.

Combining temperature and polarization data covering different sky
areas is straightforward when working in pixel space, an important
feature due to the different Galactic cuts possible for these two data
sets.  Taking them into account in the harmonic space requires either
analytical approximations, which do not suite large--angle
experiments well (see, e.g., Ng \& Liu \cite{ng}), or extensive
Monte--Carlo simulations to calibrate suitable window functions.

In this work we perform a likelihood analysis of \textit{double
reionization} models of the Cen (\cite{cen}) type. The reionization
history is described by two parameters: besides $\tau $, which is
regarded as the reionization parameter more directly constrained by
experiments, we consider $z_r$, the redshift at which the IGM ionizes
for the first time. A second reionization is then assumed to occur at
$z=7$. The duration of the first ionized period is fixed once by
assuming that, between the two ionized eras, $x_e=1/3$. This category
of reionization histories is meant to approach the pattern indicated
by Cen (\cite{cen}). We consider a grid of models spanning a large
portion of the parameter space, and take into account temperature,
polarization and temperature--polarization cross--correlation spectra.

All of our analyses have been performed considering the features of
the SPOrt\footnote{http://sport.bo.iasf.cnr.it} experiment, including
its sky coverage which avoids the Celestial polar caps.  In addition,
we have simulated higher sensitivities in order to set experiment
requirements to significantly measure both relevant parameters, taking
into account Cosmic Variance (CV).

Cosmic reionization has its main impact on the harmonics linked to
angular scales subtending the cosmological horizon at reionization.
Accordingly, a knowledge of spectra above $l \sim 30$ has a modest
impact, so that the effect of using an angular resolution $\sim 5\degr
$--$7\degr$ is also modest. The resolution of SPOrt ($\sim 7\degr$)
allows then a fair inspection of polarization features related to this
epoch (patchiness effects, altering $l \ga 1500$ multipoles, are not
considered here). A wide beam implies that beam smoothing must be
taken into account in data analysis (on the contrary, the low--$l$
multipoles are almost free of beam effects in high--resolution
experiments). Here we verify that an accurate treatment implies no
serious limitation in reconstructing the history of physical events,
in a $7\degr$ experiment.

Similar analyses have been performed in previous works using different
reionization histories. Kaplinghat et al. (\cite{kap}) modeled a
two--step reionization, while Holder et al. (\cite{bias}) simulated
reionization histories based on physical models of formation and
evolution of the first ionizing sources. At variance with the double
reionization model by Cen (\cite{cen}) and with models considered
here, almost all models discussed in these works display a monotonic
increase of the ionized fraction with time.

All of these analyses coherently imply that, with levels of
sensitivity similar to WMAP, information on reionization going beyond
the integral $\tau $ value is hardly obtainable. An important finding
of Holder et al. (\cite{bias}) is that $\tau $ estimates resulting
from fitting a sharp reionization history to models characterized by
more complex reionization, are affected by a bias. Their numerical
analysis shows that, at the WMAP sensitivity level, the estimate of
$\tau $ lays within 1--$\sigma $ from the actual value. However,
they show that the bias grows with sensitivity and the discrepancy
between the actual and the estimated values can even go beyond $\sim
10\,\sigma $ for CV--limited experiments.

This behavior is found also for the double--reionization patterns
considered in this work. We also find that, for the noise range
discussed here, the $\tau $ estimate shifts to \textit{ greater}
values as sensitivity improves. In particular, at sensitivity $\sim$
WMAP, the optical depth tends to be underestimated, while a decrease
of noise by a factor $\sim 10$ leads to an overestimated $\tau $, well
outside 3 $\sigma $, if a sharp reionization is assumed. We also
provide an interpretation of why this trend occurs, and suggest that a
test of the shape of the reionization pattern can be performed, in
long term experiments, by comparing $\tau $ estimates performed on
data obtained in shorter or longer periods.

The paper is organized as follows: in section~\ref{sec:reio} we
provide more details on the reionization histories that we considered
and describe how the publically available code CMBFAST was modified to
allow power spectrum computation for such histories. In
section~\ref{sec:like} we describe the likelihood analysis. In
section~\ref{sec:res} we discuss the results of the analysis and in
section~\ref{sec:concl} we draw our conclusions.

\section{Reionization pattern and $E$--mode spectrum} 
\label{sec:reio} 

When the Universe reionizes, the CMB photon distribution has a
relevant quadrupole term, greatly enhanced with respect to the last
scattering epoch. This enables Thomson scattering polarization to be
preserved in the distribution of scattered photons. Accordingly, the
CMB polarization rate (CMBP) depends on the scattered fraction and,
therefore, on $\tau$. The polarization distribution on spherical
harmonics, instead, depends on the evolution of the ionized fraction,
so that the polarization spectrum bears an imprint of the reionization
history (see, e.g., Zaldarriaga \cite{zald97}, Kaplinghat et
al. \cite{kap}, Holder et al. \cite{bias}, Naselsky \& Chiang
\cite{nachiang}). A quantitative evaluation of these effects can only
be done numerically. Here we report results obtained by suitably
modifying the publically available linear code CMBFAST (Seljak \&
Zaldarriaga \cite{selza96}).

\begin{figure}      
\begin{center}  
\includegraphics[width=1\hsize]{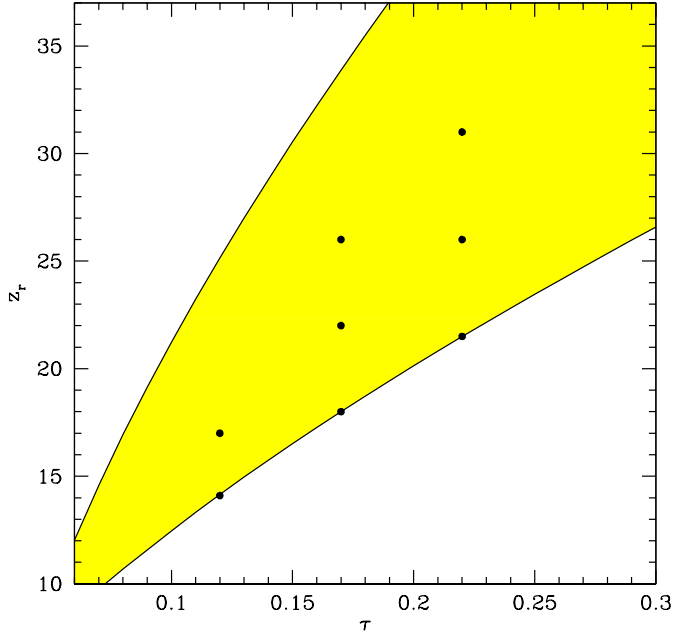}	
\caption{The parameter space allowed in our investigations in the plane
$\tau$--vs--$z_r$. Dots show the position of fiducial models analyzed in 
this work. Fiducial models falling on the bottom edge have a single 
reionization.} 
\label{fig:area}
\vskip -.8truecm
\end{center}  
\end{figure}

We select a flat $\Lambda$CDM cosmology with Hubble parameter (in
units of 100~km/s/Mpc) $h = 0.71$ and density parameters $\Omega_m h^2
= 0.148$, $\Omega_b h^2 = 0.024$.  We then take a grid of points in
the $\tau$--$z_r$ parameter plane, spanning the intervals $0.07 < \tau
< 0.30$ and $10 < z_r < 39$ (see Figure~\ref{fig:area}).

Double reionization allows the same value of $\tau$ with different
$z_r$.  The minimal allowed $z_r$ for such a $\tau$ value is obtained
for a single sharp reionization occurring at a suitable $z$; this sets
the lower side of the shaded area in Figure~\ref{fig:area}.
Then, the low--ionization period, lasting from a redshift $z_1$ to
$z_2 = 7$, reduces to zero. For any greater $z_r$,
\begin{equation}
\tau \simeq \tau_2 +{ \rho_b(z_2) \over m_p } 
\sigma_T c t_2 \left[ \big( {1 +z_r \over 1 + z_2} \big)^{3 \over 2} -
{2 \over 3} \big( {1 +z_1 \over 1 +z_2} \big)^{3 \over 2} - {1 \over 3} \right]
\label{eq:z1}
\end{equation}
Here $\sigma_T$ is the Thomson cross--section, $m_p$ is the average
baryon mass, $\rho_b(z_2)$ the baryon matter density at redshift
$z_2$; $t_2$ is the time corresponding to $z_2$, $\tau_{2}$ is the
optical depth due to full reionization since $t_2$. In
eq.~(\ref{eq:z1}) a matter dominated expansion is assumed, since $z_r$
to $z_2=7$, yielding an error of a few percent. Given $\tau$ and the
$(1 +z_r)/(1 +z_2)$ ratio, eq.~(\ref{eq:z1}) fixes the ratio $(1
+z_1)/(1 +z_2)$. There is however a top value for $z_r$, achieved when
$z_r = z_1$; this sets the upper side of the shaded area in
Figure~\ref{fig:area} (obtained through exact numerical
integration). Then, the first $x_e = 1$ period, lasting from $z_r$ to
$z_1$, reduces to zero.

A first reionization occurring at a redshift $z_r \gg 30$ is unlikely
under most models of ionizing sources (see, e.g., Haiman \& Holder
\cite{haho}); therefore, not all points falling within the shaded area
of Fig.~\ref{fig:area} bear the same physical relevance, in spite of
being compatible with the parametrization of reionization 
discussed here.

\begin{figure}      
\begin{center}  
\includegraphics[width=1\hsize]{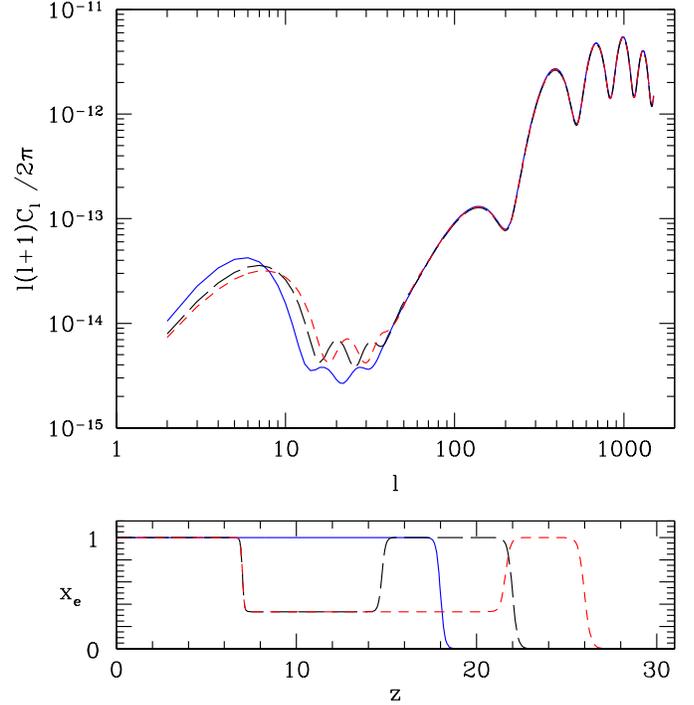}	
\caption{E-mode angular power spectra (top panel) and ionization
history (bottom panel) for three different models considered in our
analysis with $\tau = 0.17$. Solid line represent a ionization profile
with $z_r = 18$, long--dashed line has $z_r = 22$ and short--dashed
line $z_r = 26$.  Different ionization levels are connected by a fast
but smooth transition in order to guarantee stability in numerical
integration.}
\label{fig:sametau}	
\end{center}  
\end{figure}			  
\begin{figure}      
\begin{center}  
\includegraphics[width=1\hsize]{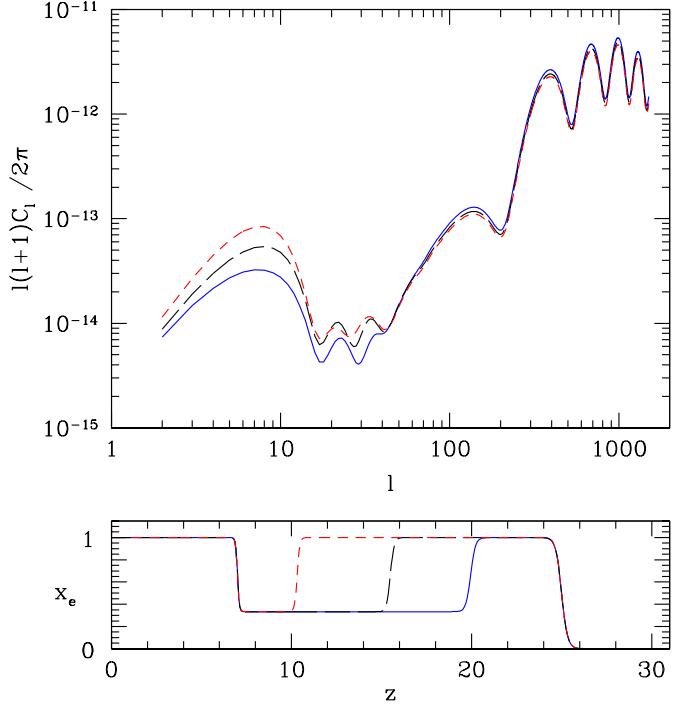}	
\caption{$C_l^E$ power spectra and ionization histories for three
models with $z_r = 25$ and $\tau =0.17$ (solid line), $\tau =0.21$
(long-dashed line), $\tau =0.25$ (short-dashed line).}
\label{fig:samezr}	
\end{center}  
\end{figure}			  

When modifying the CMBFAST linear code, we are very careful in
treating on ionization transients. Analytical and numerical
approximations at each shift are the same as used in CMBFAST when
dealing with a single reionization event. In particular, to avoid
instabilities in numerical integration, due to fast variations in
$x_e$, we adopt grid steps $s_\tau =.01$ and $s_z =1$.

In Figure~\ref{fig:sametau}, we show the varying ionization rates and
the resulting CMBP $E$--mode spectra for a set of models with equal
$\tau$ but different $z_r$. Conversely, in Figure~\ref{fig:samezr}, we
show models where the first reionization redshift $z_r$ is kept
constant and $\tau$ is variable. The models of
Figure~\ref{fig:sametau} lie on a vertical line of
Figure~\ref{fig:area}, those of Figure~\ref{fig:samezr} lie on a
horizontal line in the same figure.  Power spectra are clearly
sensitive to the ionization history and potentially encode information
on it. Notice that varying $\tau$ has effects extending to greater
$l$, while varying $z_r$ produces effects limited to $l \la 50$.  In
the next sections we discuss how far these variations are
detectable at different levels of instrumental sensitivity.

\section{Likelihood Analysis} 
\label{sec:like} 

Large angular scales are significantly affected by CV and a given
model can yield significantly different skies. To evaluate
how far large angle CMB experiments can recover the
reionization history, we adopt a Monte Carlo approach.

The basic outline of our approach is as follows. We select a fiducial
cosmological model $\tilde{\cal M}$ and generate 5000 sky maps, with
resolution and beam smoothing similar to those of actual
experiments. To each CMB map we add a noise map; the resulting maps
for the anisotropy $T$ and the $Q$, $U$ Stokes parameters are then
used for likelihood evaluation. The probability distribution function
(p.d.f.) for the parameters characterizing the model $\tilde{\cal M}$
is then simply the distribution of the best--fit parameters in the
likelihood analysis of such maps.

In this work we deal with polarization data similar to those expected
from the SPOrt experiment, however allowing for a higher sensitivity,
while temperature data are supposed to come from an experiment similar
to WMAP. Simulated maps are generated using the
HEALPix\footnote{http://www.eso.org/science/healpix/} package. Both
$T$ and $Q, U$ data are smoothed with a $7\degr$ FWHM Gaussian filter
and we chose a HEALPix resolution $N_{side} =16$ (corresponding to a
pixel width of $\sim 3.5\degr$). This approach results in a rather
low number of pixels, while parameter extraction is sensitive only to
$l \la 40$ multipoles. Each simulated map represents a random
realization of the process $\tilde{\cal M} +{\cal N}$. The noise model
${\cal N}$ is assumed to be uniform and white, and is then fully
defined by the rms noise values $\sigma_T$ and $\sigma_P$ for $T$ and
$Q,~U$ pixels.

As explained in the Introduction, we remove the region with Galactic
latitude $|b| < 20\degr$ from $T$ maps, where Galactic contamination
strongly surpasses the CMB signal. On the other hand, the synchrotron
template developed by Bernardi et al. (2004) shows that at 90~GHz the
synchrotron polarized emission in the Galactic Plane is at most
comparable with the CMB polarization signal for a cosmological model
with optical depth $\tau = 0.17$. Taking into account that foreground
removing techniques allow one to lower the foreground contamination by
a significant amount (by a factor $\sim 5$--10), the residual
contamination is likely to affect the CMB polarization analysis to a
negligible extent, and we choose not to remove the Galactic Plane in
the present analysis of ($Q,~U$) maps. Still, part of the Galactic
Plane is cut out by excluding declinations $|\delta| > 51.6\degr$,
which SPOrt is unable to inspect. (In addition, we notice that
polarized emission by dust grains is expected to lie safely below
synchrotron, at least up to $l \simeq 50$; see Fabbri \cite{fabbri}
for a more detailed discussion).

Simulated data are ordered into a vector ${\bf x} \equiv
\{T(i=1,...,N_T), Q(i=1,....,N_P), U(i=1,....,N_P)\}$, $N_T$ and $N_P$
being respectively the number of anisotropy and polarization
pixels. The likelihood of a model ${\cal M}$ is then given by a
multivariate Gaussian:
\begin{equation}
{\cal L}({\cal M}|{\bf x}) = {1 \over (2\pi)^{N_T+2N_P}} 
{1 \over \sqrt{\det{\bf C}}}
\exp \left(- {1 \over 2}{\bf x}^{\rm T}{\bf C}^{-1}{\bf x}\right) ~;
\label{eq:like}
\end{equation}
 where the correlation matrix reads:
\begin{equation}
{\bf C}_{ij}  \equiv \langle {\bf x}^T_i{\bf x}_j\rangle 
={\bf S}_{ij}+{\bf N}_{ij}
\label{eq:veco}
\end{equation} 
(brackets mean ensemble average). Its elements represent the expected
correlation between elements of the vector ${\bf x}$. For example, the
expected correlation between $T$ signals in the $i$ and $j$ pixels, at
an angular distance $\vartheta_{ij}$, reads
\begin{equation}
\langle T_i T_j \rangle = \sum_l {2l + 1 \over 4\pi} C^T_l P_l (\cos
\vartheta_{ij}) B^2_l + (\sigma_T)^2 \delta_{ij};
\end{equation}
here $C^T_l$ is the anisotropy spectrum of the model, $P_l (\cos
\theta)$ are Legendre polynomials, while the coefficients $B^2_l$
account for pixelization and beam smoothing. For a FWHM of $7\degr$,
the correction $B^2_l$ is relevant even for the lowest harmonics.
Similar expressions, taking into account the tensor nature of
polarization, hold for correlations involving $Q$ and $U$ data (see,
e.g., Zaldarriaga \cite{zald98}, Ng \& Liu \cite{ng}). The effects of
sky cuts, here, are directly set by the $i, j$ index domains. This is
a clear advantage of working in the coordinate space.  (As already
outlined, in the harmonic space, $f_{sky}$ corrections are unsafe for
large angle experiments or when $T$ and $Q, U$ data cover different
portions of the sky.)

For each sky realization, we seek the values of the reionization
parameters which maximize the likelihood function $\cal L$. Repeating
the operation for a set of realizations allows us to study the
resulting distributions.

The fiducial models considered in this work are shown in Figure
\ref{fig:area}. Each model was tested at 3 levels of sensitivity for
the polarization measures: $\sigma_P = 1.50,~0.45,~0.15 \,\mu$K for
$\sim 7\degr$ pixels, corresponding to $\sim 10, \, 3, \, 1
\,\mu$K--degree. The first value is the expected sensitivity for the
90GHz SPOrt channel. The pixel noise for $T$ data is set at the
reference value $\sigma_T = 1 \mu$K.  For such $\sigma_T$,
uncertainties on measurement of $T$--multipoles are dominated by the
CV up to $l \approx 500$ while, with our beamwidth and pixelization,
only the first $\la 40$-50 multipoles matter. Thus, a $\sigma_T$
reduction would not yield a significant gain. On the contrary, as
polarization is $\sim 100$ times smaller than anisotropy, sensitivity
increases like those considered here improve the ratio between CV and
noise variance, mainly in the harmonic range we are exploring.  How
the different $\sigma_P$ values interfere with polarization multipoles
must be considered in detail, to understand our results, and will be
discussed below.

The distribution of the best fit parameters, among the 5000 CMB
realizations of each fiducial model, at the three noise levels,
provides a frequentist estimate of the probability density function
(p.d.f.) of model parameters. To ensure that CV did not introduce
spurious effects in the comparisons of different models, we used the
same set of random seeds and of noise maps for all fiducial
models. Therefore: (i) Differences between sets of sky maps with the
same noise level are due only to changes in the CMB power spectra
between the fiducial models. (ii) Differences between sets of maps at
different sensitivities (but corresponding to the same model) are only
due to variations in the noise.

\section{Results} 
\label{sec:res} 
All fiducial models were analyzed first assuming a single
reionization, then considering reionization histories of the kind
described in Sec.~\ref{sec:reio}. We compare the probability
distributions on $\tau$, to test the bias induced by the {\it prior}
of sharp reionization.

\begin{table}
\caption{Fiducial models: $\tau$ is the total optical depth, $z_r$ is
the redshift of the first reionization. Models with the superscript
$s$ have a sharp reionization history, other models assume that the
Universe is completely reionized at $z <7$, while between these two
reionization periods $x_e$ drops to 1/3.}
\label{tab:mods}
\vglue0.2truecm

\begin{center}
\begin{tabular}{ccccccccc}
\hline
\rule[-1ex]{0pt}{3.5ex}
          &
 {${\cal A}^s$} &  
 {$\cal B$} &  
 {${\cal C}^s$} &  
 {$\cal D$} &  
 {$\cal E$} &
 {${\cal F}^s$} &
 {$\cal G$} &
 {$\cal H$} 
  \\
\hline
$\tau$  &  0.12 &  0.12  &  0.17  &  0.17  &  0.17  &  0.22  &  0.22  &  0.22  
\\
$z_r$   &   14.1  &  17  &  18.0  &  22   &  26   & 21.4 &  26   &  31  
\\
\hline
\end{tabular}
\end{center}
\end{table}

\begin{figure}      
\begin{center}  
\includegraphics[width=1\hsize]{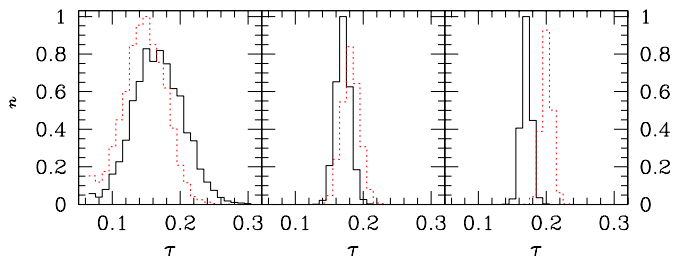}	
\caption{Distribution of $\tau$ for models ${\cal D}$, at different
noise sensitivity (left to right, $\sigma_P = 1.50,\, 0.45,\, 0.15 \,
\mu$K). Solid lines show results obtained marginalizing over $z_r$,
dotted lines refer to results of checking the same realizations
against single reionization models. At high sensitivity, the latter
prior give rise to a noticeable bias.}
\label{fig:tbias}	
\end{center}  
\end{figure}			  
\begin{figure}      
\resizebox{\hsize}{!}{\includegraphics{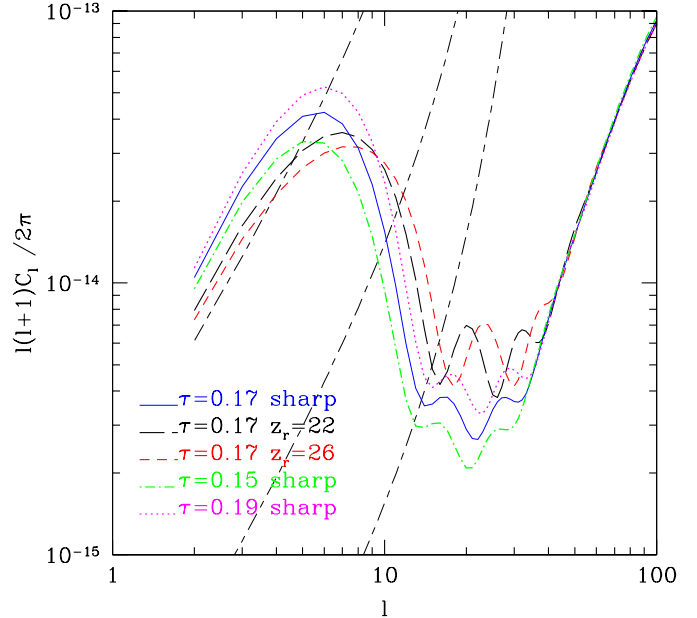}}
\caption{E--mode angular power spectra for models ${\cal C}$ (solid
lines), ${\cal D}$ (long--dashed lines), ${\cal E}$ (short--dashed)
and two sharp reionization models with $\tau = 0.15$, 0.19
(dot--dashed and dotted lines respectively). At given value of the
optical depth, spectra of models with double reionization fall below
the APS of a sharp reionization with same $\tau $ for $ l ~{\tilde <}
~8$--10 and above for $10  ~{\tilde <} ~l ~{\tilde <} ~40$. The
long--short dashed lines, instead, show the noise spectra for
different pixel noises on polarization (top to bottom $\sigma_P =
1.50, \, 0.45, \, 0.15 \, \mu$K), corrected for pixelization and beam
smoothing.}
\label{fig:clbias}	
\end{figure}			  

In Fig.~\ref{fig:tbias}, we compare the $\tau$ distributions obtained
under the prior of single reionization (dotted lines) with those
obtained after marginalizing over $z_r$ the joint distribution on
$\tau$ and $z_r$ (solid lines). The three plots account for three
polarization noise levels. At each level, no significant difference
between the width of the two distributions is apparent. However, while
all marginalized distributions peak at the {\it true} optical depth
$\tau =0.17$, the single reionization distributions show a bias,
depending on $\sigma_P$, similarly to Holder et al. (\cite{bias}).  Let
us however outline a further trend: if $\sigma_P =1.50 \,\mu$K, the
single reionization prior leads to underestimating $\tau$, although
within the (still wide) statistical error; as $\sigma_P$ decreases,
the peak $\tau $--value increases and, at the top sensitivity
considered, $\tau$ is overestimated by more than two standard
deviations.

This happens because the range of $C^E_l$ significantly affecting the
estimates is strongly related to $\sigma_P$, for the sensitivity
levels we are considering. In fact, $\tau$ accounts for the number of
scattered CMB photons, which can affect the total amount of large
scale polarization (Zal\-dar\-ria\-ga \cite{zald97}). Therefore, the
$E$--spectra, for models with different $\tau$ but equal $z_r$, differ
mainly in the height of the first reionization peak. But a variation
of $z_r$ --~at fixed $\tau$~-- also shifts the peak in
$l$. Fig. \ref{fig:clbias} shows the $E$--spectra for the models
${\cal C, D, E}$, plus two {\it single reionization} models, with
$\tau = 0.15$ and $\tau = 0.19$. For models with $\tau = 0.17$, when
$z_r$ increases, the first reionization peak moves to greater $l$ and
its height slightly decreases: then $l \la 8$ harmonics of models
${\cal D}$ and ${\cal E}$ approach those for single reionization with
lower $\tau$; the same models, in the range $8 \la l \la 20$, resemble
a single reionization with greater $\tau$.

When $\sigma_P = 1.50\, \mu$K (top noise level considered),
reionization parameters are mostly determined by the first 6--7
multipoles, the only ones clearly above noise (see
Fig.~\ref{fig:clbias}). Trying to fit a single reionization to a
double reionization history therefore leads to underestimating $\tau$.
As the noise decreases and multipoles in the range $10 \la l \la 20$
acquire a greater weight, a double reionization history can be
misinterpreted as single reionization with a greater $\tau$.

The shift of the reionization peak to greater $l$ is typical of models
with complex reionization histories. Hence, this low--$\sigma$
high--$\tau$ effect, when a single reionization is assumed, should be
independent of reionization details, for the range of sensitivities
considered here. For higher sensitivities, when multipoles $20 \la l
\la 40$ affect the $\tau$ determination, the $E$--spectrum exhibits
further peaks, whose details are related to model features, and the
trend will depend on the set of models considered.

In general, the bias is stronger for greater $\tau$ and, at fixed
$\tau$, it increases with $z_r$. At noise levels accessible to current
experiments' statistical errors exceed the bias, and $\tau$ estimates
are safe, although systematically smaller than the real values. In
higher sensitivity experiments, wrong priors can cause a $\tau$
overestimate, exceeding several standard deviations. An opposite
behavior is found when single reionization models are analyzed
assuming a two--reionization history. In this case, however, the bias
is not so severe.

The low--$\sigma$ high--$\tau$ effect can be used to test the
reionization pattern, by comparing early and late outputs, in a long
lived experiment. Detecting a clear trend requires a sensitivity
increase by at least a factor $\ga 3$. As we are considering low
multipoles, early outputs must cover the whole sky with a reasonable
sensitivity. Assuming that a full sky coverage requires $\sim 6$
months, the total experimental lifetime should be $\sim 5$--6 years.
The required increase is unlikely within the planned lifetime of
current experiments, but could be an option for future polarization
missions.

We performed some more detailed tests of this point.  (i) For each
realization we labeled $\tau^s_{1.50}, \, \tau^s_{0.45} \,
\tau^s_{0.15}$ ($\tau^d_{1.50}, \, \tau^d_{0.45} \, \tau^d_{0.15}$)
the optical depth estimated at $\sigma_P = 1.5,~0.45,~0.15\, \mu$K,
under a single (double) reionization prior. (ii) We selected the
realizations where the $\tau$ estimates, under a single reionization
prior, increased at both noise reductions ($\tau^s_{1.50} <
\tau^s_{0.45} < \tau^s_{0.15}$). (iii) Among them we kept those for
which the shift $\tau^s_{0.15} -\tau^s_{1.50}$ exceeded twice
$\tau^d_{0.45} -\tau^d_{1.50}$. This residual fraction averages $\sim
65 \%$; although varying from $\sim 20 \%$ (model ${\cal B}$) to more
than $90 \%$ (model ${\cal H}$). A progressive shift to higher values
in single--reionization $\tau$ estimates, when sensitivity increases,
therefore can be considered as a hint that a more refined
description of reionization is needed.

\begin{figure}      
\begin{center}  
\includegraphics[width=1\hsize]{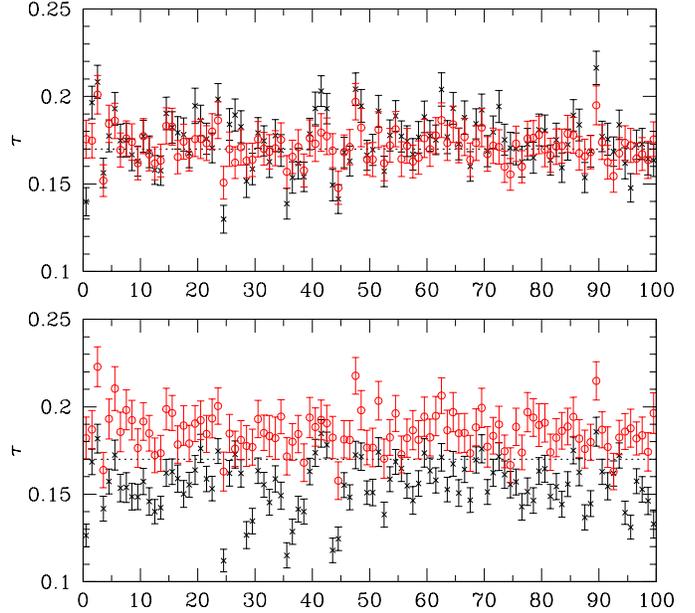}	
\caption{Top. Open circles: $\tau$ from
a WMAP+SPOrt--like experiment, with $\sigma_P = 0.40 \mu$K
in 5 years, for 100 realizations of model ${\cal D}$, and with a fair
double reionization prior. Crosses: average $\tau$ obtained by
dividing the whole data set into 10 subsets (6 months observation time). 
Error bars yield the 1--$\sigma$ c.l.  
Bottom. The same as Top panel but assuming a sharp reionization prior.
Global (averaged) estimates are shifted up(down)--ward in respect to the
actual $\tau$ value.}
\label{fig:medie}	
\end{center}  
\end{figure}			  
\begin{figure}      
\begin{center}  
\includegraphics[width=0.8\hsize]{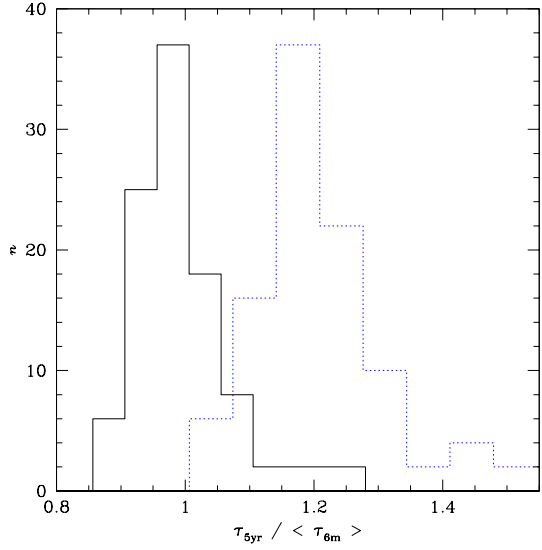}	
\caption{Distribution of the ratio between global and averaged $\tau$
estimates in 100 realizations of model ${\cal D}$, assuming either
double realizations (solid line) or sharp reionization (dotted line) priors.}
\label{fig:ratio}	
\end{center}  
\end{figure}			  

An additional test, for long lived experiments, is provided by
dividing the global dataset into subsets corresponding to shorter
observation periods, and comparing the $\tau$ value estimated by the
analysis of the full dataset with the weighted average of the
estimates in the smaller subsets. For reference purposes, we
considered a SPOrt--like experiment lasting 5 years, and achieving a
polarization pixel sensitivity $\sigma_P = 0.40 \mu$K ($N_{side} =
16$). Neglecting correlated noise effects and assuming that a full sky
coverage requires 6 months of observations, the global data can be
divided into 10 smaller subsets, characterized by $\sigma_P \sim 1.25
\mu$K. We then compared the global $\tau$ estimate with the average of
the estimates in the smaller subsets. Repeating this analysis for 100
random realizations of model $\cal D$, we found global and averaged
estimates to be consistent with each other and with the actual $\tau$,
when correct priors are made (top panel in Fig.~\ref{fig:medie}). On
the contrary, when sharp reionization is assumed, averaged estimates
are systematically lower than the correct value, while global
measurements overestimate it (bottom panel in Fig.~\ref{fig:medie}).
Then, in Fig.~\ref{fig:ratio}, we plot the distribution of the ratio
between the full 5--year estimate and the averaged 6--month
estimates. Assuming a correct prior, such ratio averages $\tau_{5 yr}
/ \langle\tau_{6m}\rangle \, = 0.99 \pm 0.07$, while for sharp
reionization $\tau_{5 yr} / \langle\tau_{6m}\rangle \, = 1.20 \pm
0.10$.

We now adopt a double reionization prior, and discuss the precision by
which $\tau$ and $z_r$ can be recovered. Marginalizing the joint
probability distribution in the $\tau -z_r$ plane on either parameter
provides the 1D probability density for the other parameter. The
variances of these distributions, averaged over fiducial models, tell
us how far the models can be discriminated at each sensitivity;
results are displayed in Table \ref{tab:sdev}: two equal--$\tau$
models can be distinguished if their $z_r$ are farther apart than
$\sim$twice the value shown. These uncertainties are close to those
obtained in the analysis of the performance of high resolution
experiments with equivalent sensitivities (see, e.g., Kaplinghat et
al.~\cite{kap}).  Thus, low angular resolution is not a serious
impediment in reionization studies.

\begin{table}
\caption{Average statistical errors for the recovered parameters as a
function of polarization sensitivity $\sigma_P \, (\mu$K).}
\label{tab:sdev}
\vglue0.2truecm
\begin{center}
\begin{tabular}{ccc}
\hline
\rule[-1ex]{0pt}{3.5ex}
 {$\sigma_P$} &  
 {$\Delta_\tau$} &  
 {$\Delta_{z_r}$} \\
\hline
 1.50 & 0.037 & 5.0 \\
 0.45 & 0.012 & 2.6 \\
 0.15 & 0.008 &  1.4 \\
\hline
\end{tabular}
\end{center}
\end{table}

Within the allowed region of parameter space (see Fig.~\ref{fig:area}
and discussion), Table~\ref{tab:sdev} implies that $\tau$ is better
fixed than $z_r$, if the priors on the reionization history are correct.
This confirms that also for double reionization, optical depth is the
most relevant parameter. At a fixed noise level, models with higher
$\tau$ generally allow for a better estimation of both parameters, due
to their higher signal-to-noise (S/N) ratio; among models with the
same optical depth those with lower $z_r$ have slightly sharper
redshift distributions, because for very early first reionization,
the differences between spectra with distinct $z_r$ are less pronounced.

\begin{figure}      
\begin{center}  
\includegraphics[width=1\hsize]{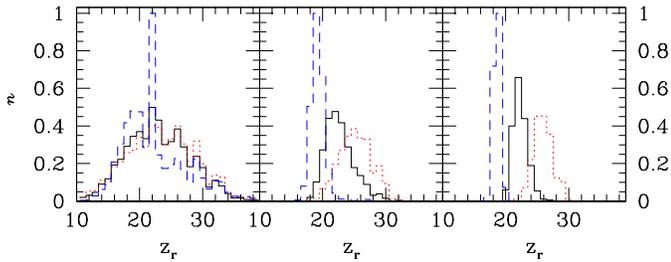}	
\caption{Distribution of maximum likelihood $z_r$, after marginalizing over 
$\tau$, for 5000 realizations of models ${\cal C}$ (dashed lines), 
${\cal D}$ (solid lines) and ${\cal E}$ (dotted lines). From left to 
right, panels refer to  $\sigma_P = 1.50,\, 0.45,\, 0.15 \, \mu$K, 
respectively.}
\label{fig:zr_t17}	
\end{center}  
\end{figure}			  
\begin{figure}      
\begin{center}  
\includegraphics[width=1\hsize]{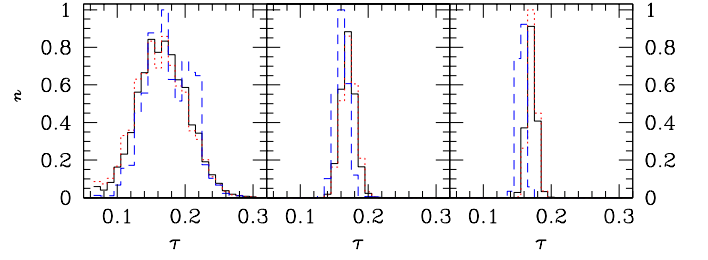}	
\caption{Same as Fig.~\ref{fig:zr_t17}, but for $\tau$ after 
marginalization over $z_r$.}
\label{fig:tau_t17}	
\end{center}  
\end{figure}			  
\begin{figure}      
\begin{center}  
\includegraphics[width=1\hsize]{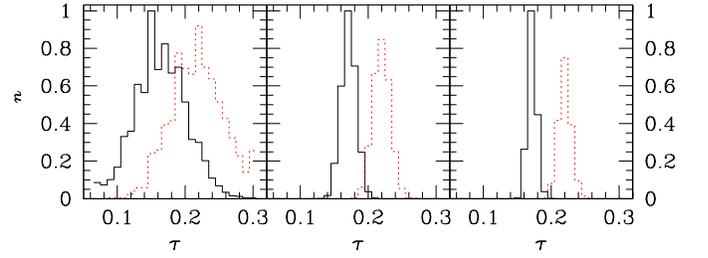}	
\caption{Same as Fig. \ref{fig:tau_t17}, but for models ${\cal E}$ 
(solid lines) and ${\cal G}$ (dotted lines).}
\label{fig:tau_z26}	
\end{center}  
\end{figure}			  
\begin{figure}      
\begin{center}  
\includegraphics[width=1\hsize]{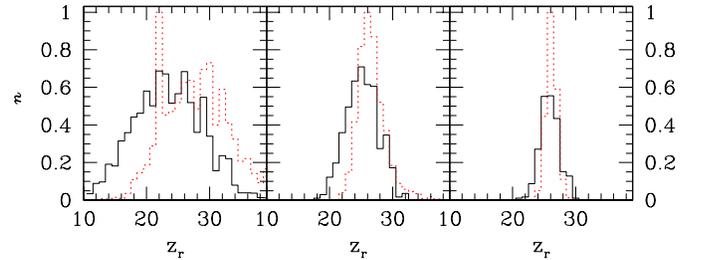}	
\caption{Same as Fig. \ref{fig:tau_z26}, but for $z_r$.}
\label{fig:zr_z26}	
\end{center}  
\end{figure}			  

In more detail, Fig.~\ref{fig:zr_t17} shows the distribution of $z_r$
for models with $\tau = 0.17$ (see caption for details). For $\sigma_P
= 1.50 \mu$K (left panel), the probability distributions overlap
significantly, and $z_r$ cannot be recovered, while for $\sigma_P =
0.15 \mu$K (right panel) they are almost completely distinguished. The
middle panel displays an intermediate situation where models with
single reionization can be distinguished from double reionization
models with $z_r$ in the upper half of the allowed range (i.e. $z_r
\ga 25$ for $\tau = 0.17$).  Moreover, distributions on $z_r$, in
double reionization models, are wider by a factor $\sim 2$ than in
single reionization models. Figure~\ref{fig:tau_t17} shows that the
probability distributions on $\tau$, in models ${\cal D}$ and ${\cal
E}$, are indistinguishable at all sensitivities; thus, differences in
$z_r$ do not greatly affect the ability to recover $\tau$, if correct
priors are used. Model ${\cal C}$, instead, displays a bias, as
discussed above. In this case, $\tau$ is progressively underestimated
for decreasing $\sigma_P$, as a single reionization model is analyzed
assuming a double reionization prior.

In Fig.~\ref{fig:tau_z26} we plot the distributions on $\tau$ for
models with equal $z_r$ but different $\tau$ (${\cal E}$ and ${\cal
G}$). Even at the highest noise, a difference can be seen; for
$\sigma_P= 0.45 \,\mu$K the two distributions are clearly
separated. In both cases, the correct $\tau$ is recovered.
Figure~\ref{fig:zr_z26}, instead, shows results for $z_r$. The
probability distributions of both models exhibit a significant overlap
and correctly peak at the true value of the reionization
redshift. Model ${\cal G}$, however, is characterized by a sharper and
better defined distribution, due to its higher S/N ratio.

We can summarize our findings as follows. (i) Constraints on the
reionization history from large--angle experiments are similar to
those by high--resolution experiments with equivalent sensitivity. For
instance, a polarization noise $\sigma_P = 1.50 \, \mu$K allows one to
constrain the total optical depth with an accuracy of $\sim 20 \%$ for
$\tau = 0.17$. Increasing sensitivity by an order of magnitude allows
for a simultaneous detection of $\tau$ and $z_r$, with an accuracy
$\la 5\%$. (ii) A wrong prior on the reionization history causes a
bias in the $\tau$ estimate. For $\sigma_P = 1.50 \, \mu$K the
statistical uncertainty exceeds the bias, while for $\sigma_P = 0.15
\, \mu$K the bias may exceed 3 standard deviations. This point, too,
agrees with previous works studying the capabilities of high
resolution CMB polarization measurements. Thus high resolution, by
itself, does not provide more reliable estimates. (iii) At current
noise levels, $\tau$ is underestimated, while, for increasing
sensitivities, it tends to be overestimated. This is likely due to the
first reionization peak moving to higher $l$'s in models with complex
reionization, with respect to sharp reionization models with the same
$\tau$.

(iv) Finally, we suggest an observational approach to discriminate
between single and double reionization histories, based on such
low--$\sigma$ high--$\tau$ effect. For illustration, we
considered a mission lasting $5\, $years, allowing a pixel
sensitivity $\sigma_P \sim 0.4\, \mu$K. We assumed that full sky
coverage requires 6 months so that the full dataset can be divided
into 10 smaller subsets, with a sensitivity worse by a factor $\sim
3$. Assuming no systematic effects, the 10 subsets provide as many
independent $\tau$ estimates. When the correct priors on reionization
are assumed, the global estimate is consistent with the average of the
estimates from the smaller subsets; on the contrary, assuming a sharp
reionization prior, the 5--year $\tau$ value systematically exceeds
both the correct value and the average of the 6--month estimates.
Accordingly, in future low--noise experiments, it can be significant
to compare whole--run $\tau$ estimates, with full sensitivity, with
averages among short run estimates. Similar error bars are expected, but
a shift of the estimated $\tau$ would indicate that the assumed
reionization pattern could be a source of bias.

\section{Conclusions}
\label{sec:concl}

In this work we discussed a class of physically motivated double
reionization models, characterized by two parameters: the total
optical depth $\tau$, and the first reionization redshift $z_r$.  We
determined at which sensitivity level their features can be recovered
by large angle CMBP measurements. We find that wrong priors on the
history of reionization can lead to a bias in $\tau$ estimates, in
agreement with findings for high resolution experiments (Holder et
al. \cite{bias}). At the WMAP or SPOrt noise level, the bias is well
within statistical errors; at higher sensitivities the $\tau$ estimate
can lie several standard deviations from its true value. Holder et
al. (\cite{bias}) argue that fitting a two--step reionization to the
models they considered allows one to significantly reduce the bias. In
addition we find that, for the class of models considered here, the
biased $\tau$ estimates exhibit a characteristic dependence on
experimental sensitivity. Testing this low--$\sigma$ high--$\tau$
effect in actual experiments provides an indication of the fairness of
priors.

Within the context of double reionization models, a pixel noise
$\sigma_P =1.50 \, \mu$K for $7\degr$ pixels, allows one to constrain
$\tau$ with an accuracy $\sim 20 \%$; an increase in sensitivity by a
factor $\sim 3$ enables us to distinguish single reionization models
from models with an early reionization (i.e. $z_r \ga 25$ for $\tau
=0.17$) followed by a partial recombination period.  Models of
ionizing sources rarely yield $z_r \ga 30$.  Firm measurements of both
$\tau$ and $z_r$, with precisions $\sim 5 \%$, require a sensitivity
increase by an order of magnitude.  Greater spatial resolution,
instead, is not very relevant, as most information on reionization is
carried by the first 30--40 multipoles of the $E$--spectrum.

Measures of $C_l^E$ at large angular scales are an important probe of
the evolution of ionizing sources at redshifts $10 \la z \la
30$. However, as finer data become available, the situation becomes
more and more risky. Unless an accurate class of models is fitted to
data, parameter misestimates can occur: apparent errors can seem
small, while true values lie well off the 3 $\sigma$ error interval. A
test of the reliability of the overall $\tau$ estimate can be however
performed by comparing estimates from the whole data set with those
from smaller representative subsets. A clear detection of this
low--$\sigma$ high--$\tau$ effect requires an increase in sensitivity
by a factor $\sim 3$. Assuming that full sky coverage requires 6
months, the effect can be evident in experiments lasting $\sim 5$--6
years.

\begin{acknowledgements}
This work has been carried out as part of the SPOrt experiment, a
programme funded by ASI (Italian Space Agency).
Some of the results of this paper were obtained with the CMBFAST and 
HEALPix packages. 
\end{acknowledgements}

\end{document}